# Клоны MV2 алгоритма

*Copyright 2000 © Лаврёнов А.*

*E-mail: lanin99@mail.ru*

Abstract: The clones of MV2 algorithm for any radix are discussed. The three various examples of ones are represented.

В работе [1] был предложен алгоритм универсального сжатия (MV2 алгоритм), основанный на побитовой (двоичной) перекодировки файла. Из-за ее неоднозначной реализации возможно существование клонов указанного алгоритма. Поэтому целью данной работы является их анализ с учетом обобщения процедуры перекодировки на другие системы исчисления. Вначале дадим несколько определений:

**Пит** – величина, принимающая ровно *p* значений.

**Пайт или n-разрядный кортеж** – последовательность, состоящая из *n* пит.

**Алфавит разрядности n ( $A_n$ )** – множество всех различных пайтов.

**Файл** - любая последовательность пайтов, возможно с повторениями.

**Основной файл** - последовательность всех элементов $A_n$ без повторения.

**Длина файла** – количество пайтов, используемых для представления файла.

**Питовая перекодировка** $A_n$ – взаимооднозначное отображение между множеством $A_n$ с постоянной длиной кортежа *n* и конечным множеством алфавитов $A_k$ с длиной кортежа $k \le n$.

**Коэффициент сжатия** $k = \dfrac{L_2}{L_1}$ - отношение длин основного файла после $L_2$ и до $L_1$ питовой перекодировки $A_n$.

Использование вычислительной техники диктует цифровое представление алфавита, которое основывается в настоящее время на позиционной (фактически – двоичной) системе исчисления. Но оно избыточно по двум причинам. Во-первых, общее число элементов $A_n$ предполагается равным $p^n$, где *p*- это число-основание системы исчисления, *n* – число-разрядность, или длина кортежа. Во-вторых, любой элемент алфавита имеет постоянную длину кортежа. Однако в повседневной жизни мы пишем, например, 13, а не 000013. Это наблюдение, а точнее преобразование – питовая перекодировка $A_n$ позволяет естественным образом ввести в рассмотрение такой признак как длина кортежа. Согласно ему первоначальный файл разделяется на совокупность вторичных файлов, не несущих смысловую нагрузку исходного текста. Один вторичный файл называется остатком и получен из исходного в результате преобразования, аналогичного вышеупомянутому для *p=2*, а второй файл – флаг несет в себе информацию об используемой длине кортежа для каждого элемента. Благодаря очевидному сжатию первоначального файла по сравнению с остатком, вся процедура может быть повторена несколько раз. Это и есть все ключевые моменты MV2 алгоритма.

Дальнейшее рассмотрение клонов обсуждаемого алгоритма будет проводиться в *p=q+1*-системе исчисления, если не оговорено специально. Для лучшего понимания и отображения сути дела на схемах реализации клонов будут представлены на примере основного файла в двоичной системе исчисления. Итак, ниже на схеме №1 наглядно показана реализация первого клона. Цифровая форма любого элемента алфавита $A_n$, расположенного в крайнем левом столбике, показана определенным образом. Незначащие нули специально выделены полужирным начертанием. Отбрасывая их, мы получим остаток, который показан крайним правым столбиком. Два средних столбика образуют нам флаг. Столбик из единиц необходим для обозначения конца кортежа. Оставшийся столбик дает информацию о количестве отброшенных нулях, т. е. длине используемого кортежа.

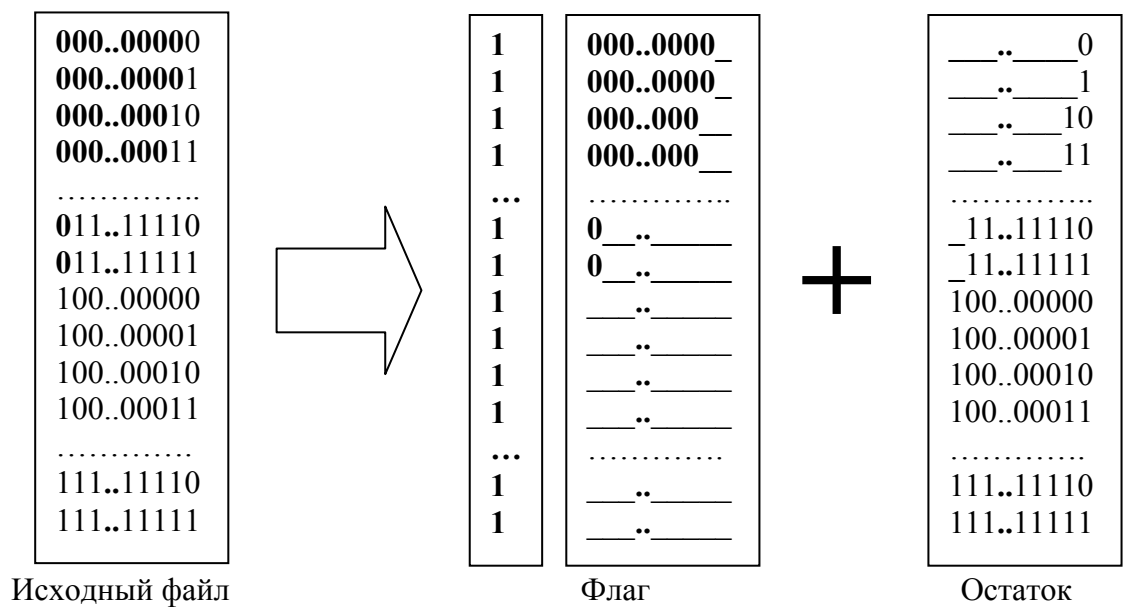

Исходный файл        Флаг        Остаток

Схема №1. Реализация первого клона MV2 алгоритма
в двоичной системе исчисления (*p=2*).

Как нетрудно заметить для остатка в схеме №1, мы фактически преобразуем только половину (в общем случае - $\frac{1}{p}$-ую часть) всех элементов. Будем иметь следующие оценки для коэффициента сжатия

$$k_1 = \frac{np^n - (n+1)p^{n-1} + 1}{n(p-1)p^{n-1}}; \text{ [для } \textbf{\textit{p=2, n=8}} \ k_1 = \frac{897}{1024} \approx 0{,}876\,]$$

для длины флага

$$L_f = \frac{p^{n+1} - p}{p - 1}. \text{[для } \textbf{\textit{p=2, n=8}} \ L_f = 510\,]$$

Очевидно, что общая длина всех вторичных файлов (остатка $L_y$ и флага $L_f$) увеличивается по сравнению с длиной первоначального файла $L_1$ на $\Delta L = p^n$, т. е. в $k_f = \frac{n+1}{n}$ раз. Выполняя эту процедуру *m*-раз, получим для клона №1 следующую величину увеличения $\frac{(k_f - 1)k_1^m + k_1 - k_f}{k_1 - 1}$ [для *p=2, n=8, m=1( 10)* имеем 1,125(1,7)].

Далее существуют два качественно разных направлений по созданию клонов. Одно из них (тип (а)) связано с размножением уже использованного набора элементов алфавитов $A_{k \leq n}$, а второе (тип (б)) – с другим набором (в пределе - с полным множеством всех элементов алфавитов $A_{k \leq n}$). Первое направление требует ввод дополнительных признаков. Поэтому мы специально остановились на показе основного файла и на возрастающей последовательности цифрового представления элементов $A_n$ на схеме №1.

Нетрудно заметить, что для всех разрядов, кроме старшего, ситуация повторяется. Это позволяет реализовать клон типа (а) при введении такого дополнительного признака как значение старшего разряда для каждого элемента. В этом случае мы получаем третий дополнительный, также не несущий смысловой нагрузки исходного текста файл - флаг №2. Наглядно данная реализация клона типа (а) показана на схеме №2.

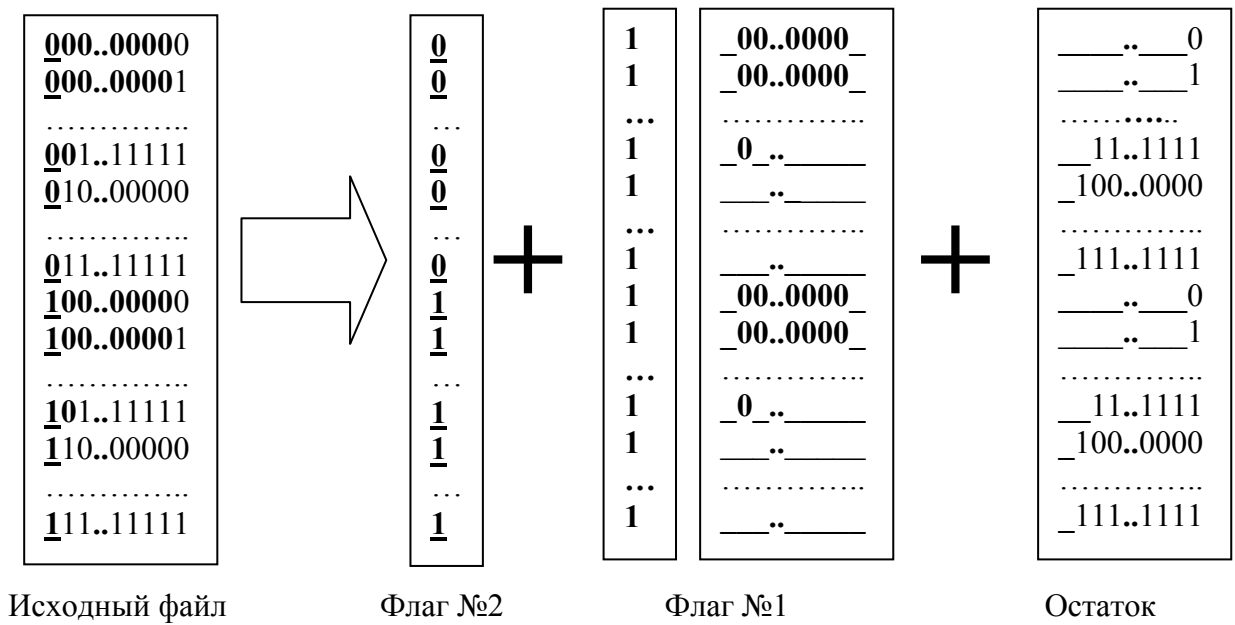

Схема №2. Реализация второго клона MV2 алгоритма
в двоичной системе исчисления (***p=2***).

В данном случае в остатке преобразуются уже все элементы $A_n$. Имеем следующие оценки для коэффициента сжатия

$$k_2 = \frac{(n-1)p^{n-1} - np^{n-2} + 1}{n(p-1)p^{n-2}}; [\text{для } \boldsymbol{p=2}, \boldsymbol{n=8} \;\; k_2 = \frac{384}{512} \approx 0{,}75]$$

для длин флагов №1 и №2 соответственно

$$L_{f1} = p^n \text{ и } L_{f2} = \frac{p^{n+2} - p^2}{p-1} \; [\text{для } \boldsymbol{p=2}, \boldsymbol{n=8} \;\; L_{f1} + L_{f2} = 256 + 1020].$$

Очевидно, что общая длина всех вторичных файлов (остатка $L_2$ и флагов $L_{f1}, L_{f2}$) та же как и для клона №1, т. е. увеличивается по сравнению с длиной первоначального файла $L_1$ в $k_f = \dfrac{n+1}{n}$ раз. Выполняя эту процедуру **m**-раз, получим для клона №2 следующую величину увеличения $\dfrac{(k_f - 1)k_2^m + k_2 - k_f}{k_{21} - 1}$ [для **p=2, n=8, m=1( 10)** имеем 1,125(1,47)].

Реализация клона типа (б) наглядно представлена на схеме №3 при наиболее полном использовании множества алфавитов $A_k$ с длиной кортежа $k \leq n$. В этом случае мы воспроизводим результаты [1]. Обобщая их результаты на **p=q+1**-систему исчисления [2], получим для коэффициента сжатия

$$k_3 = \dfrac{np^{n+2} - p^{n+1}(2n+1) + p^n n + p^2 n + p(1-n)}{np^n(p-1)^2}; [для\ \textbf{p=2, n=8}\ k_2 = \dfrac{777}{1024} \approx 0{,}759\ ]$$

для длины флага а) *p=2*
$$L_f = 3*2^n - 2n - 2\ [для\ \textbf{p=2, n=8}\ L_f = 750\ ];$$

б) *p≠2*, $S_m = \dfrac{p^m - 1}{p - 1} \geq n - 1$

$$L_f = p^n + \dfrac{(m-1)(p-1)^{m+1} - m(p-1)^m + p - 1}{(p-2)^2}.$$

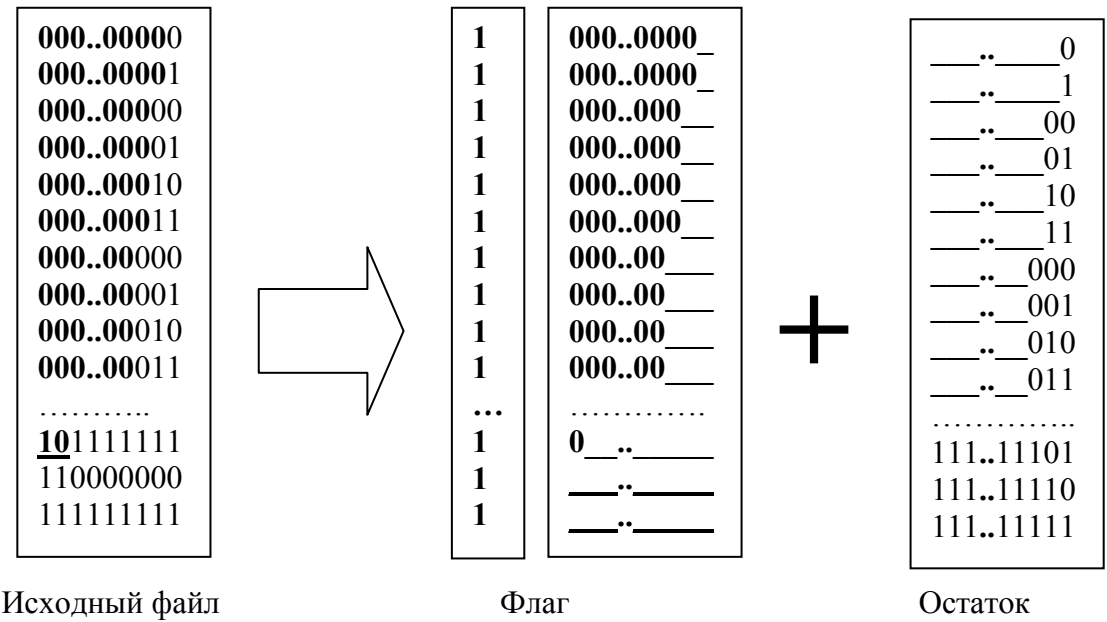

Схема №3. Реализация MV2 алгоритма по работе [1] (***p=2***).

Таким образом, рассмотрено три качественно разных клонов MV2 алгоритма с естественным выбором наборов элементов алфавитов $A_{k \leq n}$ для питовой перекодировки. Комбинируя их или вводя свои собственные, возможно реализовывать другие клоны. Однако ввод признаков для такого набора элементов алфавитов $A_{k \leq n}$ для питовой перекодировки требует отдельного обоснования. В заключение, интересно отметить, что для *p=2, n=8* мы имеем наибольший коэффициент сжатия для клона №2. Результаты статьи докладывались на конференции "Наука и педагогика на рубеже 21 столетия", посвященной 10-летию образования ИСЗ (Беларусь, Минск, 26-27 октября 2000 г.)

Литература

[1]. Виланский Ю.В., Мищенко В.А. Кодирование информации на основе алгоритма универсального сжатия. Вести ИСЗ 2000 г., №1, стр. 36-39.
[2]. Лаврёнов А. Теоретический предел сжатия информации
http://xxx.lanl.gov/abs/cs.cr/0208002